# Eliminating the wavefunction from quantum dynamics: the bi-Hamilton-Jacobi theory, trajectories and time reversal

Peter Holland[†]

**Abstract**

We observe that Schrödinger's equation may be written as two real coupled Hamilton-Jacobi (HJ)-like equations, each involving a quantum potential. Developing our established programme of representing the quantum state through exact free-standing deterministic trajectory models, it is shown how quantum evolution may be treated as the autonomous propagation of two coupled congruences. The wavefunction at a point is derived from two action functions, each generated by a single trajectory. The model shows that conservation as expressed through a continuity equation is not a necessary component of a trajectory theory of state. Probability is determined by the difference in the action functions, not by the congruence densities. The theory also illustrates how time-reversal symmetry may be implemented through the collective behaviour of elements that individually disobey the conventional transformation (T) of displacement (scalar) and velocity (reversal). We prove that an integral curve of the linear superposition of two vectors can be derived algebraically from the integral curves of one of the constituent vectors labelled by integral curves associated with the other constituent. A corollary establishes relations between displacement functions in diverse trajectory models, including where the functions obey different symmetry transformations. This is illustrated by showing that a (T-obeying) de Broglie-Bohm trajectory is a sequence of points on the (non-T) HJ trajectories, and vice versa.

## 1 Introduction. Alternatives to the wavefunction

In a series of works, the author has presented ways in which quantum evolution may be represented exactly using trajectories (nondenumerable ensembles of moving points) as state variables in place of the wavefunction. The Schrödinger equation becomes a set of autonomous deterministic equations governing the trajectories, the evolution of the wavefunction version of state being deduced from their collective flow. Two classes of trajectory models have been investigated, distinguished by how they connect with the wavefunction when it is expressed in terms of real functions: using polar variables ($\psi = \sqrt{\rho} \exp(iS/\hbar)$) ([1-3] and references in [4]), or real and imaginary parts ($\psi = \psi_R + i\psi_I$) [5-7]. The wavefunction at a spacetime point is built from either a single path propagating data from a unique initial space point (polar model) or two paths emanating from two initial points (real/imaginary model). These methods make essential use of the continuity equation whose general solution employs trajectories to link the initial and present densities. Borrowing terminology from continuum physics, the relation between the wave and trajectory varieties of state corresponds to that between the Eulerian and Lagrangian pictures [8], for single- or two-phase fluids. To set the new constructive trajectory technique proposed in this article in context, we first elaborate on the two existing schemes.

---

[†] Green Templeton College, University of Oxford, Oxford, UK. peter.holland@gtc.ox.ac.uk

Using the polar representation, the Schrödinger equation

$$i\hbar \frac{\partial \psi}{\partial t} = -\frac{\hbar^2}{2m}\partial_{ii}\psi + V(x)\psi \qquad (1.1)$$

becomes the two real equations

$$\frac{\partial \rho}{\partial t} + \frac{1}{m}\partial_i(\rho \partial_i S) = 0 \qquad (1.2)$$

$$\frac{\partial S}{\partial t} + \frac{1}{2m}\partial_i S \partial_i S + Q + V = 0 \qquad (1.3)$$

where $Q(x,t) = -(\hbar^2/2m\sqrt{\rho})\partial_{ii}\sqrt{\rho}$ is the quantum potential ($i,j,... = 1,2,3$). This formulation suggests the well-known hydrodynamic interpretation [9] in which the quantum state is conceived in terms of the density $\rho$ and velocity potential $S$ of a putative quantum fluid. An alternative sobriquet for (1.3) is 'the quantum Hamilton-Jacobi (HJ) equation', since it is redolent of the classical HJ equation if $mv^i = \partial_i S$ is identified with momentum, so that $(\partial_i S)^2/2m$ is a kinetic energy, and $Q$ is regarded as a potential energy additional to the external potential $V$. Note, though, that this is not an HJ equation in the classical sense because it does not stem from a transformation theory [10], and the mutual dependence of the $\rho$ and $S$ fields means that the 'external' potential $Q$ is $S$-dependent and tacitly involves derivatives of $S$ higher than the first (example: for a one-dimensional stationary state with a finite current, $Q = -\hbar^2(\partial S)^{1/2}\partial^2(\partial S)^{-1/2}/2m$) [9]). The latter property implies that the separation into kinetic and potential energies used in (1.3) is arbitrary and, as we show below, other hydrodynamic- or HJ-like renderings of (1.1) are possible, involving alternative identifications of these energies as functions of the wavefunction.

The dynamics of the exact self-contained trajectory theory of state corresponding to the polar representation is governed by a second-order differential equation, or equivalently an integro-differential first-order equation [1-4]. We denote its solutions, which coincide with the trajectories of the de Broglie-Bohm interpretation[1], by $q^i(q_0,t)$ where $q_0^i$ uniquely labels a path and may be taken to be the initial location. This specification of the state is supplemented by the initial density $\rho_0$ and velocity $\dot{q}_0^i = m^{-1}\partial_i S_0$. The time-dependent wavefunction is derived from the trajectories, up to a global phase, thus:

$$\psi(x,t) = \sqrt{\rho_0(q_0)J^{-1}(q_0,t)}\,\exp[i\chi(q_0,t)/\hbar]\big|_{q_0(x,t)} \qquad (1.4)$$

with

---

[1] It is emphasized that the trajectory constructions of state described in this paper are alternative mathematical formulations of quantum theory and are neutral as regards interpretation. In particular, in the polar theory no path is singled out as physically special or adorned with a substantive corpuscle, the central assumption of the de Broglie-Bohm theory. These models do, however, have the potential to expand the scope of interpretations, as we show below in this section for the conventional interpretation.

$$\chi(q_0, t) = \int m\dot{q}^i \frac{\partial q^i}{\partial q_0^j} dq_0^j + L(q(q_0, t), t) dt \tag{1.5}$$

where $\dot{q}^i = \partial q^i(q_0, t)/\partial t|_{q_0}$, $L(q(q_0, t), t) = \frac{1}{2} m\dot{q}^{i\,2} - Q(q) - V(q)$, $J = \det(\partial q/\partial q_0)$ and $\rho = \rho_0 J^{-1}$ is substituted in $Q$. The integral (1.5) is taken along any line joining the points $(q_0^i = 0, t = 0)$ and $(q_0^i, t)$. The amplitude $\psi$ at a spacetime point $(x^i, t)$ is thereby built from a single trajectory propagating $\psi_0$ from the unique initial point $q_0^i$ (the notation $q_0(x, t)$ means that $q_0^i$ is chosen so that $q^i(q_0, t) = x^i$). Note, though, that a path does not subsist in isolation, as in a particle-mechanical system, but contributes to a nondenumerable ensemble of paths, analogous in this regard to an optical ray. Moreover, the assembly is a *congruence* (so there are no focal points), and neighbouring trajectories mutually interact through attractions and repulsions whose specifically quantum character is embodied in $Q$ (which depends on the $q_0^i$-derivatives of $q^i$ and $\rho_0$). This is the causal mechanism driving quantum evolution in this model.

The trajectory picture undermines 'wavefunction hegemony' [11] in conceiving the quantum state but the two viewpoints are complementary rather than antagonistic. Indeed, each view is implicit in the other: when wave mechanics is formulated in terms of (gauge) potentials for the Eulerian functions $\rho$ and $v^i$, the two pictures are connected by a canonical transformation [1,4]. A consequence is that the trajectory provides an additional perspective on the physical meaning of the wavefunction, consistent with its usual (Eulerian) interpretation.

We shall examine this point briefly by considering the Lagrangian interpretations of the Eulerian functions $\rho$ and $S$. The conventional interpretation of $\rho$ is that $\rho d^3 x$ is the probability of finding the position in the domain $(x^i, x^i + dx^i)$ at time $t$. If the measurement of position is formulated in Lagrangian terms (which requires extending the trajectory formalism to a many-body system, as in [3]) the outcome at an instant $t$ is one of the possible current locations $q^i(q_0, t)$, distinguished by the label $q_0^i$. Since the trajectory through a spacetime point is unique, we may infer the label $q_0^i(q, t)$ from the outcome and thus treat the process as a measurement of label. Hence, we obtain two interpretations of the distribution in measurement results: uncertainty in the position (usual Eulerian view) or in the label of a trajectory (Lagrangian view). It is easy to see that these interpretations are indistinguishable using only position measurements. Thus, in the Lagrangian picture, the local conservation of probability corresponding to (1.2) reads

$$\frac{\partial}{\partial t}[\rho(q, t) d^3 q(t)]\bigg|_{q_0} = \frac{\partial}{\partial t}(PJ d^3 q_0)\bigg|_{q_0} = 0 \tag{1.6}$$

where $P(q_0, t) = \rho_0(q_0) J^{-1}(q_0, t)$. The probability of finding the trajectory label in the range $(q_0^i, q_0^i + dq_0^i)$, $PJ d^3 q_0$, is therefore equal to $\rho d^3 x$ for all $t$ and so the two interpretations are empirically equivalent.[2]

---

[2] The measurement of the trajectory velocity is discussed in [4].

Turning to the phase $S$, its Lagrangian-picture representative $\chi$ in (1.5) is an action function, specifically the action of a particle of mass $m$ starting from the point $q_0^i$ and moving in a given potential $Q + V$.[3] Two useful forms for the action may be obtained from the general expression (1.5) by choosing integration paths $(A)$ $(0,0) \to (q_0^i, 0) \to (q_0^i, t)$ or $(B)$ $(0,0) \to (0, t) \to (q_0^i, t)$:

$$\chi(q_0, t) \stackrel{A}{=} S_0(q_0) + \int_0^t L(q(q_0, t), t) dt \bigg|_{q_0} \tag{1.7}$$

$$\stackrel{B}{=} f(t) + \int_0^{q_0^j} m\dot{q}^i \frac{\partial q^i}{\partial q_0^j} dq_0^j \bigg|_t \tag{1.8}$$

where $f(t) = \int_0^t L(q(q_0, t), t)|_{q_0=0} dt$. To express the action in the Eulerian picture, we make the change of variables $q_0^i \to q_0^i(x, t)$ in (1.5) and, defining $p_i = mv^i(x, t) = m\dot{q}^i(q_0(x, t), t)$ and $H(x, p, t) = p_i^2/2m + V(x, t) + Q(x, t)$, we obtain the classical-type formula

$$S(x, t) = \int p_i dx^i - H dt. \tag{1.9}$$

When extended to Riemannian space, the polar approach provides trajectory models for a wide class of quantum and non-quantum systems. As an example, one accomplishment is the solution to a longstanding problem that appears insuperable in the wavefunction language, that of representing exactly a many-body system in terms of states defined in three-dimensional space [3].

In the second class of earlier models of state, connected with real and imaginary components, the free Schrödinger equation splits into a pair of coupled continuity equations where $\psi_R$ and $\psi_I$ play the role of conserved densities, and the two associated velocity fields are functions of them [5]. In the associated exact stand-alone trajectory theory, the state is defined by two vector functions $q_R^i(q_{R0}, t)$ and $q_I^i(q_{I0}, t)$ alongside the initial densities $\psi_{R0}$ and $\psi_{I0}$, and is governed by first-order coupled differential equations. The trajectories propagate the system according to the formula

$$\psi(x, t) = \psi_{R0}(q_{R0}) J_R^{-1}(q_{R0}, t)|_{q_{R0}(x,t)} + i\psi_{I0}(q_{I0}) J_I^{-1}(q_{I0}, t)|_{q_{I0}(x,t)}. \tag{1.10}$$

In this case, the amplitude $\psi$ at a point is built from two trajectories propagating the initial values $\psi_{R0}$ and $\psi_{I0}$ that originate from the points $q_{R0}^i$ and $q_{I0}^i$, respectively. The formula (1.10) may be generalized to embrace external fields [5], and the method extends to trajectory constructions of solutions to the wave and Klein-Gordon equations [6], and the Dirac equation [7]. A notable feature of these models is that the spacetime boost symmetry is implemented by non-classical transformations of the Eulerian velocity fields, in both the non-relativistic [5] and relativistic [7] cases. In the latter case it has been shown that the corresponding non-standard Lagrangian transformations must be label dependent to achieve Lorentz covariance [7,12].

---

[3] The action for the congruence is given by $\int \rho_0(q_0) \left[\frac{1}{2}m\dot{q}^{i2} - (\hbar^2/8m)(\partial\log\rho/\partial q^i)^2 - V(q(q_0))\right] d^3q_0 dt$. A trajectory is subject to a force $-\frac{\partial V}{\partial q^i} - \frac{1}{\rho}\frac{\partial \sigma_{ij}}{\partial q^j}$ where $\sigma_{ij} = -\frac{\hbar^2 \rho}{4m}\frac{\partial^2 \log\rho}{\partial q^i \partial q^j}$. Its coincidence with the path of a particle of mass $m$ in the potential $V + Q$ follows from the identity $\frac{1}{\rho}\frac{\partial \sigma_{ij}}{\partial q^j} = \frac{\partial Q}{\partial q^i}$ [1].

The first method described above employs a continuity equation (1.2) and an HJ-like equation (1.3) as field equations, and the second uses two continuity equations. In this contribution we complete the triptych of possible representations of Schrödinger's equation that employ continuity and HJ-like equations by showing how it can be expressed as a pair of real coupled HJ-like equations, each involving a quantum potential (§2). We show (§3) that this formulation also admits an associated exact trajectory construction comprising the autonomous propagation of two coupled congruences, which respectively generate action functions that together represent the wavefunction. The bi-HJ-like picture shares some features with the two preceding models but exhibits significant differences. It shows that conservation as expressed through a continuity equation obeyed by functions of the wavefunction is not a necessary component of a trajectory theory of state; probability is determined here by the difference in the action functions. It also illustrates how time-reversal symmetry may be implemented by transformations that disobey the conventional transformation (T) of displacement components (scalars) and velocity (reversal); rather, the time reversal transform of one flow is the T-reversal of the other (§§2 and 3). We are then led to examine how the trajectories in the various models we have proposed are connected. In §4.1, we prove that an integral curve of the linear superposition of two vectors can be derived algebraically from the integral curves of one of the constituent vectors labelled by integral curves associated with the other constituent. A corollary establishes relations between the displacement functions in different trajectory models, including where the functions obey different symmetry transformations. In §4.2, we employ this construction to show how a conserved flow may be built from a non-conserved flow and, applying the theory to the two HJ-like flows, prove that the evolution of probability cannot be generated by their trajectory densities. The corollary is illustrated in §5 by showing that a (T-obeying) de Broglie-Bohm trajectory is a sequence of points on the (non-T) HJ-like (and other similar) trajectories, and vice versa.

The theory was prompted by a recent unified theory of particle and wave [13,14] in which time-reversal covariance plays a significant role. However, as already noted[1], the work herein concerns just an alternative formulation of the quantum formalism, without the additional physical elements of the unified model.

## 2 Schrödinger's equation as a bi-Hamilton-Jacobi-like system

We shall represent the Eulerian version of the quantum state, i.e., $\psi$, using the two real functions $S_\pm = S \pm (\hbar/2)\log\rho$.[4] In terms of them, the wavefunction decomposes into a product of two complex amplitudes,

$$\psi = e^{(1+i)S_+/2\hbar}e^{(-1+i)S_-/2\hbar}, \tag{2.1}$$

which individually do not obey Schrödinger's equation. The latter becomes two real coupled equations, closed in the fields $S_\pm$:

$$\frac{\partial S_+}{\partial t} + \frac{1}{2m}\partial_i S_+ \partial_i S_+ + Q_+ + V = 0 \tag{2.2}$$

$$\frac{\partial S_-}{\partial t} + \frac{1}{2m}\partial_i S_- \partial_i S_- + Q_- + V = 0 \tag{2.3}$$

where

$$Q_\pm = \mp \frac{\hbar}{2m}\partial_{ii} S_\mp - \frac{1}{4m}[\partial_i(S_+ - S_-)]^2. \tag{2.4}$$

This formulation must be supplemented by conditions corresponding to those obeyed by $\psi$. For example, single-valuedness is embodied in the constraints $\oint \partial_i S_\pm dx^i = nh, n \in \mathbb{Z}$. An external vector potential may be included in (2.2)-(2.4) by replacing $\partial_i S_\pm \to \partial_i S_\pm - A^i$ but we shall not need this.

We obtain in (2.2) and (2.3) what looks like a pair of Hamilton-Jacobi equations if we treat the functions $S_\pm$ as potentials for two velocity fields,

$$v_\pm^i(x,t) = m^{-1}\partial_i S_\pm, \tag{2.5}$$

and regard the terms $Q_\pm$, which couple the equations, as 'quantum potentials', nonclassical additions to the classical HJ equation. The functions $Q_\pm$ share the distinctive properties of $Q$ in (1.3) [9]: they are second order in the coordinates, gauge invariant, dependent on the form of $\psi$ rather than its absolute magnitude, and they do not generally fall off with distance. Following our remarks above regarding the arbitrariness in apportioning energy in (1.3), this representation offers alternative identifications of potential and kinetic energies as functions of the wavefunction. However, as with (1.3), (2.2) and (2.3) are not HJ equations in the classical sense: they have not been derived from a transformation theory, the quantum potentials depend (explicitly here) on the velocities and their derivatives, and, of course, there are two coupled equations. Nevertheless, as we shall see, there is sufficient commonality in the classical and quantum formalisms to warrant use of the 'HJ' epithet. Henceforth, we shall often write 'HJ' for 'HJ-like'.

Taking the gradients of (2.2) and (2.3) gives coupled acceleration equations:

$$\frac{\partial v_\pm^i}{\partial t} + v_\pm^j \partial_j v_\pm^i = -\frac{1}{m}\partial_i\left[\mp\frac{\hbar}{2}\partial_j v_\mp^j - \frac{m}{4}\left(v_+^j - v_-^j\right)^2 + V\right]. \tag{2.6}$$

---

[4] It is tacitly assumed that the function $\log\rho$, where $\rho$ has the dimension $l^{-3}$, contains a constant that renders it dimensionless. This constant adds to the indetermination in $S$ and does not affect any results.

These equations, in conjunction with (2.5), constitute an alternative version of Schrödinger's equation ((2.2) and (2.3) are easily recovered from them). In a hydrodynamic analogy, (2.6) suggests visualizing a quantum system as a mixture of two miscible fluids that intermingle throughout space, each point supporting both flows. In the Eulerian picture of the mixture, (2.6) may be interpreted as a pair of Euler force equations governing a bi-potential flow (2.5). The scheme has two notable properties. First, adapting ideas of the continuum theory of mixtures [15,16], the coupled character of the equations, i.e., the appearance of $v_\mp^i$ in the equation for $v_\pm^i$, is suggestive of a two-phase model that exhibits continual conversion of one species of fluid into the other. Second, the theory is couched just in terms of velocity fields and is devoid of independent density functions, an uncommon occurrence in hydrodynamics. This is evident from the acceleration potentials on the right-hand side of (2.6), and from the absence of continuity equations. The individual fluids may be attributed Lagrangian trajectory densities (see §§3 and 4.2), but the corresponding Eulerian fields relate to the velocity fields, and then only indirectly through their divergences. The Eulerian probability density is also connected with the velocities, but in a quire different way via the potential difference:[5]

$$\rho = \exp[(S_+ - S_-)/\hbar]. \tag{2.7}$$

In the context of (2.2) and (2.3), the continuous symmetries of the ten-parameter Galileo covariance group of the Schrödinger equation take the form

$$\begin{aligned} t' &= t + d, \quad x'^i = a_{ij}x^j - w^i t + c^i, \\ S'_\pm(x', t') &= S_\pm(x, t) + a_{ij}w^i x^j + m\, w^{i^2} t/2, \end{aligned} \tag{2.8}$$

where $d$, $c^i$, $a_{ij}$ (with $a_{ij}a_{ik} = \delta_{jk}$) and $w^i$ are constants, and $V$ is assumed to be a scalar. It follows that each of the velocity fields (2.5) is a Galileo 3-vector ($v'^i_\pm = a_{ij}v^j_\pm - w^i$), a property they share with the single velocity vector employed in the de Broglie-Bohm theory and in classical mechanics. The formalism diverges from the latter theories, however, when we consider the time-reversal symmetry of (1.1) (again, we assume $V$ is a scalar). This transformation involves the complex conjugate wavefunction [18-20] and reads $\psi'_R = \psi_R, \psi'_I = -\psi_I$ in the real variables, or

$$\text{T:} \quad x'^i = x^i, \quad t' = -t, \quad \rho'(x', t') = \rho(x, t), \quad S'(x', t') = -S(x, t). \tag{2.9}$$

The sign reversal of the phase $S$ ensures the covariance of the HJ equation (1.3) and entails the reversal of the de Broglie-Bohm velocity: $v'^i = -v^i$. This is in accord with time reversal in classical mechanics, where the HJ function reverses. We denote by T this 'standard' transformation. The corresponding transformations of the new fields are

$$S'_{\pm}(x',t') = -S_{\mp}(x,t) \tag{2.10}$$

whence

$$v'^i_{\pm}(x',t') = -v^i_{\mp}(x,t). \tag{2.11}$$

---

[5] The relation (2.7) evokes Boltzmann's formula connecting probability and entropy ($P = e^{S_B/k}$), an observation that might become significant should a connection be established between action and entropy (as suggested by de Broglie [17]).

It is readily checked that the functions $S'_{\pm}$ are solutions of (2.2) and (2.3) (in the primed variables). Evidently, in this view the velocities characterizing the two flows do not exhibit the usual sign reversal property (as do the de Broglie-Bohm and classical velocities); rather, they map into the negative of each other. The theory is effectively T-covariant, but due to the collective behaviour of elements that individually disobey T. Such exchange of roles is a signature of the time-reversal covariance of (1.1), which transforms into its complex conjugate and conversely [18]. Likewise, (2.2) and (2.3) interchange, as do the Euler equations (2.6). We shall develop this point in §3.

The formalism we have introduced overlaps with that employed in the stochastic interpretation, which embellishes the quantum formalism with a random component in the velocity [21,22]. Within that scheme, the velocities (2.5) are interpreted as forward and backward drift velocities, $u^i = v^i_+ - v^i_- = (\hbar/m)\partial_i \log\rho$ is the osmotic velocity, the local mean $v^i = \frac{1}{2}(v^i_+ + v^i_-)$ is the de Broglie-Bohm velocity, and equations (4.17) below are the corresponding Fokker-Planck equations. In this connection, we stress that we are engaged here in finding an alternative expression for the deterministically evolving quantum state, and we do not introduce an additional stochastic (or any other) mechanism to modify the formalism. Our proposal to write the Schrödinger equation as two HJ equations, and to connect the velocities $v^i_{\pm}$ with a deterministic trajectory model (which we do next), does not appear to have been explored before.

Our approach should also be distinguished from the so-called 'quantum Hamilton-Jacobi' theory in which $\psi = e^{iW/\hbar}$, $W \in \mathbb{C}$, is inserted in Schrödinger's equation to turn it into a complex HJ-like equation for the 'HJ' function $W$ [23,24] (still a second-order equation in $x^i$, and thus not of the classical type). An interpretation of that formalism using complex trajectories has been proposed in which the Schrödinger equation is modified by complexifying the real coordinates $x^i$ [25]. It is not clear that this procedure provides a consistent interpretation of quantum mechanics due to problems with defining a locally conserved probability current [25,26], but this issue is beyond the scope of our present investigation.

## 3 Eliminating the wavefunction from quantum dynamics: a bi-trajectory construction of the quantum state

Complementing the Eulerian theory of the two interacting fields $S_{\pm}$, we shall show how the time-varying quantum state may be represented independently by two coupled Lagrangian congruences $q^i_+(q_{+0},t)$ and $q^i_-(q_{-0},t)$, supplemented by the initial data $S_{\pm 0}$. Here, $q^i_{+0}$ and $q^i_{-0}$ label elements of the respective continuous ensembles of trajectories and may be chosen as the initial positions. The velocities in the two pictures are connected by the relations

$$\dot{q}^i_\pm \equiv \left.\frac{\partial q^i_\pm(q_{\pm 0}, t)}{\partial t}\right|_{q_{\pm 0}} = v^i_\pm(x = q_\pm(q_{\pm 0}, t), t). \tag{3.1}$$

Given $\psi(t)$, and hence $v^i_\pm$, the trajectories may be found by solving the differential equations (3.1). Our aim is to invert this procedure and obtain from the wave equation in the form (2.6) exact self-contained trajectory equations whose solutions imply $\psi(t)$.

Each space point simultaneously supports a trajectory of each genus. Suppose that at time $t$ the paths $q^i_{+0}$ and $q^i_{-0}$ cross the point $x^i$. The relations

$$x^i = q^i_+(q_{+0}, t) = q^i_-(q_{-0}, t), \qquad i = 1,2,3, \tag{3.2}$$

uniquely fix the choice of labels $q^i_{\pm 0}(x, t)$ that ensure this intersection. This implies a mutual dependence of the labels: given $q^i_{+0}(x, t)$, the corresponding unique $q^i_{-0} = q^i_{-0}(q_{+0}, t)$.

There are two steps to turning (2.6) into stand-alone equations to calculate the time dependence of the Lagrangian coordinates. First, relations (3.1) and (3.2) are employed to generate the following coupled second-order acceleration equations:

$$\ddot{q}^i_+ = -\frac{1}{m}\frac{\partial}{\partial q^i_+}\left[\frac{\hbar}{2}\frac{\partial \log J^{-1}_-}{\partial t} - \frac{m}{4}\left(\dot{q}^j_+ - \dot{q}^j_-\right)^2 + V(q_+, t)\right], \tag{3.3}$$

$$\ddot{q}^i_- = -\frac{1}{m}\frac{\partial}{\partial q^i_-}\left[-\frac{\hbar}{2}\frac{\partial \log J^{-1}_+}{\partial t} - \frac{m}{4}\left(\dot{q}^j_+ - \dot{q}^j_-\right)^2 + V(q_-, t)\right], \tag{3.4}$$

which are subject to the initial conditions $\dot{q}^i_{\pm 0} = m^{-1}\partial S_{\pm 0}(q_{\pm 0})/\partial q^i_{\pm 0}$. Here,

$$\frac{\partial}{\partial q^i_\pm} = J^{-1}_\pm J^j_{\pm i}\frac{\partial}{\partial q^j_{\pm 0}}, \qquad J^j_{\pm i} = \frac{\partial J_\pm}{\partial(\partial q^i_\pm/\partial q^j_{\pm 0})}, \tag{3.5}$$

$J_\pm(q_{\pm 0}, t) = \det(\partial q_\pm/\partial q_{\pm 0})$, and we have used the identities $\partial \log J_\pm/\partial t = \partial \dot{q}^i_\pm/\partial q^i_\pm$. The second step is to write these coupled equations in terms of a common set of independent variables. If we choose the latter as $q^i_{+0}$ and $t$, the dependent variables are $q^i_+(q_{+0}, t)$ and $q^i_{-0}(q_{+0}, t)$. Inverting the latter vector function, the vector function $q^i_-(q_{-0}, t)$ is obtained from (3.2) as $q^i_+(q_{+0}(q_{-0}, t), t)$. The equations (3.3) and (3.4) are transformed into the common variables using the formulas

$$\dot{q}^i_- = \dot{q}^i_+ - \frac{\partial q^i_-}{\partial q^j_{-0}}\frac{\partial q^j_{-0}}{\partial t}, \qquad \frac{\partial q^i_-}{\partial q^j_{-0}} = \frac{\partial q^i_+}{\partial q^k_{+0}}\left(\frac{\partial q^j_{-0}}{\partial q^k_{+0}}\right)^{-1}, \tag{3.6}$$

and a corresponding expression for $\ddot{q}^i_-$. Naturally, we could take $q^i_{-0}$ and $t$ as independent variables instead. The two subsidiary restrictions corresponding to the single-valuedness of $\psi$ become $\oint \dot{q}^i_\pm dq^i_\pm = nh/m$, conditions that are conserved by (3.3) and (3.4) (a quantum version of Kelvin's circulation theorem).

We assert that equations (3.3) and (3.4) with the stated initial conditions constitute *an exact free-standing trajectory formulation of Schrödinger's equation.* To demonstrate this, we shall derive from them Schrödinger's equation in the Eulerian form (2.2) and (2.3) using a method that supplies a formula for the time-dependent wavefunction in terms of the trajectories. This is a quantum version of the deduction of the classical HJ equation directly from Newton's law. It is achieved by performing a Weber transformation [1,27], which entails multiplying (3.3) and (3.4) by $\partial q_\pm^i/\partial q_{\pm 0}^j$, respectively, rearranging, and integrating with respect to $t$. The result is

$$m\dot{q}_\pm^i \frac{\partial q_\pm^i}{\partial q_{\pm 0}^j} = m\dot{q}_{\pm 0}^j + \frac{\partial}{\partial q_{\pm 0}^j} \int_0^t \left(\frac{1}{2}m\dot{q}_\pm^{i\,2} - Q_\pm(q_\pm,t) - V(q_\pm,t)\right) dt \qquad (3.7)$$

where

$$Q_\pm = \pm\frac{\hbar}{2}\frac{\partial \log J_\mp^{-1}}{\partial t} - \frac{m}{4}\left(\dot{q}_+^j - \dot{q}_-^j\right)^2 \qquad (3.8)$$

and the integrand in the $\dot{q}_+^i$ ($\dot{q}_-^i$) equation is expressed as a function of $q_{+0}^i$ ($q_{-0}^i$) and $t$. This procedure puts this trajectory version of the Schrödinger equation in coupled, integro-differential, first-order form. Inserting the initial conditions $\dot{q}_{\pm 0}^i = m^{-1}\,\partial S_{\pm 0}/\partial q_{\pm 0}^i$ on the right-hand side of (3.7) implies that the velocities are the $q_{\pm 0}^i$-gradients of some potentials $\chi_\pm(q_{\pm 0},t)$, forms they retain for all time (a version of the classic result of hydrodynamics [28]). Using these potentials, the six dynamical equations (3.7) may be written in purely differential first-order form as the eight coupled equations

$$\frac{\partial \chi_\pm}{\partial q_{\pm 0}^j} = m\dot{q}_\pm^i \frac{\partial q_\pm^i}{\partial q_{\pm 0}^j} \qquad (3.9)$$

$$\frac{\partial \chi_\pm}{\partial t} = \frac{1}{2}m\dot{q}_\pm^{i\,2} - Q_\pm(q_\pm) - V(q_\pm) \qquad (3.10)$$

with initial conditions $\chi_{\pm 0} = S_{\pm 0}(q_{\pm 0})$. We claim that the functions $\chi_\pm(q_{\pm 0}(q_\pm,t),t) = S_\pm(x = q_\pm,t)$. This is easily proved by the change of independent variables $q_{\pm 0}^i, t \to x^i, t$. Then, (3.9) is equivalent to (3.1), and (3.10) to (2.2) and (2.3). We have thus obtained, via a coordinate transformation from the trajectory formulation, the Schrödinger equation in its customary form in which the state $\psi$ is wholly decoupled from the trajectories. The sets of equations (1.2) and (1.3), (2.2) and (2.3), (2.6), (3.3) and (3.4), (3.7), or (3.9) and (3.10), are all equivalent versions of Schrödinger's equation (1.1), written using Eulerian or Lagrangian fields.

Following (2.1), relations (3.9) and (3.10) imply the following formula for the time-dependent wavefunction:

$$\psi(x,t) = \prod_{\epsilon=\pm} \exp\left[\frac{(\epsilon 1 + i)}{2\hbar}\int_{0,0}^{q_{\epsilon 0}^j,t} m\dot{q}_\epsilon^i \frac{\partial q_\epsilon^i}{\partial q_{\epsilon 0}^j} dq_{\epsilon 0}^j + L_\epsilon\bigl(q_\epsilon(q_{\epsilon 0},t)\bigr)dt\right]\Bigg|_{q_{\epsilon 0}(x,t)}, \qquad (3.11)$$

where $L_\epsilon = \frac{1}{2}m\dot{q}_\epsilon^{i\,2} - Q_\epsilon(q_\epsilon) - V(q_\epsilon)$ and $\epsilon = \pm$. We thus construct $\psi$ from two line integrals, each similar to the one appearing in the polar representation (1.4). As with the latter case, the integrals in (3.11) are action functions for a particle of mass $m$, but now moving in the potential $Q_+ + V$ or $Q_- + V$. Choosing suitable integration paths, each of the line integrals (for $\epsilon = \pm$) may be written in either of the forms (1.7) or (1.8). In addition, writing $p_{\pm i} = mv_\pm^i(x,t) = m\dot{q}_\pm^i(q_{\pm 0}(x,t),t)$ and $H_\pm = p_{\pm i}^{\,2}/2m + Q_\pm(x) + V(x)$, the HJ functions have forms similar to (1.9):[6]

$$S_\pm(x,t) = \int p_{\pm i} dx^i - H_\pm dt. \tag{3.12}$$

In these variables, the probability density (2.7) becomes

$$\rho(x,t) = \exp\left(\frac{1}{\hbar}\int (p_{+i} - p_{-i})dx^i - (H_+ - H_-)dt\right). \tag{3.13}$$

Employing the Lagrangian viewpoint, therefore, the wavefunction has been expunged from the dynamical equations and features only in the initial conditions. It is represented at each point by a brace of coupled action functions $S_\pm$, each built from a single trajectory propagating the initial value $S_{\pm 0}$ from the point $q_{\pm 0}^i$, respectively. In generating the actions, the trajectories may be said to determine the flow of probability density in accordance with formula (2.7), but they do not follow the lines of probability conservation. We shall show later that the trajectories determine probability flow in a different sense in that they generate the de Broglie-Bohm paths, which do directly reflect the probability flow (§5). Note that the expansion factors $J_\pm^{-1}$, which in part define the trajectory densities and feature in the quantum potentials (3.8), relate to Eulerian velocity divergence terms in (2.6) and are not to be conflated with the probability density (see §4.2).

Using (2.7), the stationary points $x^i(t)$ of $\rho$, its peaks and troughs, are characterized by points of coincidence of the two brands of velocity: $v_+^i(x,t) = v_-^i(x,t)$. From (3.6), the trajectories $q_{0\pm}^i(t)$ occupying these points are determined by the condition $\partial q_{-0}^i(q_{+0},t)/\partial t = 0$ along with (3.2).

To examine time-reversal symmetry in this formalism, we first note that the Lagrangian-picture implementation of the field-theoretic transformation T in (2.9) is that of a classical (and de Broglie-Bohm) displacement:

$$\text{T:} \quad q_0'^i = q_0^i, \quad t' = -t, \quad q'^i(q_0',t') = q^i(q_0,t), \tag{3.14}$$

with $\dot{q}'^i(q_0',t') = -\dot{q}^i(q_0,t)$. In contrast, the Lagrangian version of (2.10) for two paths that cross the point $(x^i,t)$ entails exchange relations for the labels and displacements:

$$q_{\pm 0}'^i = q_{\mp 0}^i, \quad t' = -t, \quad q_\pm'^i(q_{\pm 0}',t') = q_\mp^i(q_{\mp 0},t). \tag{3.15}$$

---

[6] It may be of interest to compare such bi-Hamiltonian systems with those that admit more than one symplectic structure [29].

These relations are in accord with the velocity transformations (2.11), since

$$\dot{q}'^{i}_{\pm}(q'_{\pm 0}, t')|_{q'_{\pm 0}} = \left.\frac{\partial q'^{i}_{\pm}(q_{\mp 0}, t')}{\partial t'}\right|_{q_{\mp 0}} = \left.\frac{\partial q^{i}_{\mp}(q_{\mp 0}, t)}{\partial(-t)}\right|_{q_{\mp 0}} = -\dot{q}^{i}_{\mp}(q_{\mp 0}, t). \qquad (3.16)$$

The left-hand side here is $v'^{i}_{\pm}(x' = q'_{\pm}(q'_{\pm 0}, t'), t')$ and the right-hand side is $-v^{i}_{\mp}(x = q_{\mp}(q_{\mp 0}, t), t)$, which gives (2.11), since (3.2) and (3.15) imply $x'^{i} = x^{i}$.

Evidently, the transformation of the function $q^{i}_{+}$ violates T: the time-reversed location is $q^{i}_{-}(-t)$, not $q^{i}_{+}(-t)$. Yet, since the accompanying time reversal of $q^{i}_{-}(t)$ is $q^{i}_{+}(-t)$, the net effect of the transformation applied simultaneously to both orbits is the same as that obtained for a pair of paths subject to T. The ensemble of trajectories that define the state is thus effectively T-covariant, but, as mentioned in §2, this is accomplished through the collective behaviour of elements that individually disobey T (Fig. 1). The collective covariance of each pair of equations (3.2), (3.3) and (3.4), and (3.7), is achieved through the reciprocal transformation of its members, as expected from the exchange of equations in the Eulerian picture (§2). In connecting the two pictures, we note that the complex conjugation of the Eulerian $\psi$ in (3.11) is a deduction from the Lagrangian mapping (3.15) with (3.16) (as is the conjugation of (1.4) from (3.14)).

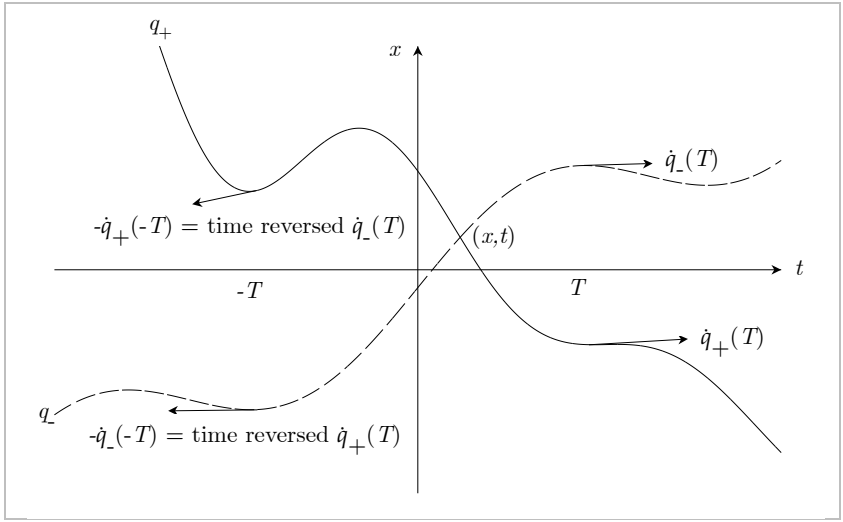

Fig. 1. The time reversal of one of the bi-HJ trajectories is the T-reversal of its partner.

The extension of the bi-trajectory theory to many-body systems is straightforward and essentially involves expanding the index range. We shall not pursue this except to observe that, for $N$ bodies, the state comprises $N$ pairs of coupled congruences in three-dimensional space, where each pair generally depends on the labels of all the pairs. The latter property is how nonlocality appears in this model. Each pair is independent of the others when the wavefunction factorizes into $N$ one-body functions, a property the model shares with the polar model [3], but not with the real/imaginary model [5].

## 4 Connecting trajectory models

### 4.1 Algebraic construction of an integral curve of a linear superposition of vector fields from integral curves associated with the constituent vectors

The trajectory model introduced in §3 relates to an Eulerian version of the quantum state obtained by a simple transformation of the dependent variables, namely, $\psi, \psi^* \to S_+, S_-$. The two other models described in §1 likewise relate to simple Eulerian reformulations (using the pairs of fields $\rho, S$ or $\psi_R, \psi_I$, which are connected by a canonical transformation [4]). In contrast, the displacement functions used in the Lagrangian coordinate models corresponding to the three Eulerian rephrasings appear to be vastly disparate with no obvious connections between them. We shall examine this problem, of expressing one species of trajectory in terms of another, with reference to the relation between the de Broglie-Bohm paths used in the wavefunction construction (1.4) and non-T paths, including the HJ tracks.

One way to obtain a de Broglie-Bohm trajectory from the paths $q_\pm^i$ is to compute the associated velocity fields via (3.1) and then derive an integral curve of the vector $v^i = \frac{1}{2}(v_+^i + v_-^i)$. Such an indirect construction does not cast any special light on the connection between the models. A second approach is to attempt to set up a canonical transformation linking the models, following the approach mentioned in §1 [4]. That idea will be examined elsewhere. We shall present here a third alternative, an algebraic method of connecting trajectory types (which is of wider scope than the quantum application made here). It follows as a corollary to the following construction in which an integral curve of the linear superposition of two vectors is derived algebraically from the integral curves of one of the constituent vectors labelled by integral curves associated with the other constituent:

**Theorem (composition of trajectories).** Let $v_A^i(x,t)$ and $v_B^i(x,t)$, $i = 1,2,3$, be two vector fields in $\mathbb{R}^3 \otimes \mathbb{R}^1$. Let $x^i = q_A^i(q_{A0}, t)$ be an integral curve of $v_A^i$ identified by the label $q_{A0}^i$: $\dot{q}_A^i = v_A^i(q_A, t)$ with $\det(\partial q_A / \partial q_{A0}) \neq 0$. Construct from $v_B^i$ the following vector function of $q_{A0}^i$,

$$V_B^i(q_{A0}, t) = \frac{\partial q_{A0}^i}{\partial q_A^j} v_B^j(q_A(q_{A0}, t), t), \qquad (4.1)$$

and let $q_{A0}^i = Q_B^i(q_{C0}, t)$ be its integral curve labelled by $q_{C0}^i$: $\dot{Q}_B^i = V_B^i(Q_B, t)$. Choose $q_{C0}^i$ so that $Q_B^i(q_{C0}, t)$ is the label $q_{A0}^i$ of the curve $q_A^i$ that occupies the point $(x^i, t)$. Then the vector function

$$q_C^i(q_{C0}, t) = q_A^i(Q_B(q_{C0}, t), t) \qquad (4.2)$$

is the integral curve through $(x^i, t)$ of the vector superposition $v_C^i = v_A^i + v_B^i$ labelled by $q_{C0}^i$: $\dot{q}_C^i = v_C^i(q_C, t)$.

**Proof.** Differentiating (4.2) with respect to $t$ implies

$$\left.\frac{\partial q_C^i}{\partial t}\right|_{q_{C0}} = \left.\frac{\partial q_A^i}{\partial t}\right|_{Q_B} + \left.\frac{\partial q_A^i}{\partial Q_B^j}\right|_t \left.\frac{\partial Q_B^j}{\partial t}\right|_{q_{C0}} \quad (4.3)$$

$$= v_A^i(q_A(Q_B(q_{C0},t),t),t) + \left.\frac{\partial q_A^i}{\partial Q_B^j}\right|_t V_B^j(Q_B(q_{C0},t),t). \quad (4.4)$$

Inverting (4.1) gives

$$v_B^i(q_A(q_{A0},t),t) = \left.\frac{\partial q_A^i}{\partial q_{A0}^j}\right|_t V_B^j(q_{A0},t). \quad (4.5)$$

Hence, evaluating (4.5) along the track $q_{A0}^i = Q_B^i(q_{C0},t)$, (4.4) may be written

$$\dot{q}_C^i(q_{C0},t) = v_A^i(q_A(Q_B(q_{C0},t),t),t) + v_B^i(q_A(Q_B(q_{C0},t),t),t) \quad (4.6)$$

$$= \left. [v_A^i(x,t) + v_B^i(x,t)] \right|_{x=q_C(q_{C0},t)} \text{ from (4.2)}. \quad (4.7)$$

Thus, $q_C^i$ is the integral curve of the vector field $v_C^i = v_A^i + v_B^i$ at the point $(x^i,t)$. ∎

**Remarks**. 1. We may regard $x^i = q_A^i(q_{A0},t)$ as a time-dependent coordinate transformation $x^i \leftrightarrow q_{A0}^i$ whereby a fixed point, attributed the coordinates $x^i$ in the space frame, is coordinated using the labels $q_{A0}^i = (q_A^i)^{-1}(x,t)$ of the succession of trajectories that pass the point as $t$ changes. Eq. (4.1) expresses the transformation of the contravariant vector $v_B^i(x,t)$ into the $q_{A0}^i$-coordinates.

2. The labels $q_{A0}^i$ and $q_{C0}^i$ identify their respective trajectories uniquely. They are defined up to time-independent diffeomorphisms and may be chosen as the respective starting points. Then $Q_B^i(t=0) = q_C^i(t=0) = q_{C0}^i$ and $q_A^i(t=0) = q_{A0}^i = f^i(q_{C0})$. At time $t$, $Q_B^i$ gives (what was) the initial position $q_{A0}^i$ of the trajectory $q_A^i$ that passes $(x^i,t)$.

3. Instead of using the integral curves of $v_A^i(q_A,t)$ and $V_B^i(q_{A0},t)$, the construction could employ the integral curves of $v_B^i(q_B,t)$ and a vector $V_A^i(q_{B0},t)$ defined analogously to (4.1).

4. The formula (4.2) may be extended to (a) $N$ space dimensions, by extending the index range, and (b) the deduction of an integral curve of the superposition of a finite number ($> 2$) of vectors from integral curves associated with the constituent vectors by applying the theorem successively to pairs of constituents.

For fixed $q_{C0}^i$, the vector function $Q_B^i$ is a 'label generator': as $t$ changes, it generates a continuous sequence of labels $q_{A0}^i = Q_B^i(t)$ each of which identifies a unique trajectory $q_A^i(q_{A0}^i)$. The successive points defined by the set of $q_A^i$s corresponding to the sequence of $q_{A0}^i$s generate the single trajectory $q_C^i(q_{C0}^i,t)$. The instantaneous direction of $q_C^i$, $\dot{q}_C^i$, is given by $[v_A^i(x,t) + v_B^i(x,t)]|_{x=q_C(q_{C0},t)}$, not $v_A^i$.

Instead of treating $v_A^i$ and $v_B^i$ as given, we may take $v_C^i$ as given and choose vectors $v_A^i$ and $v_B^i$ whose sum composes it. It follows that an integral curve of a given vector field $v_C^i$ may be ascertained algebraically from those of *any* vector $v_A^i$ defined in the same domain by choosing the complementary vector $v_B^i = v_C^i - v_A^i$ and evaluating $q_A^i(q_{A0})$ along the integral curves of the corresponding vector $V_B^i$. This could include, for example, building trajectories in one potential from those in another (an idea that has been developed previously [30]) or using vectors $v_A^i$ that are not usually regarded as velocities (such as an electric field). Of special interest here is that the vectors $q_C^i$ and $q_A^i$ may obey different transformation rules in respect of some spacetime $(x^i, t)$ symmetry for which $v_A^i$ and $v_C^i$ transform contrarily. We have:

**Corollary**. An integral curve of a vector field $v_C^i$ may be expressed as a sequence of points lying on a set of integral curves of any vector field $v_A^i$ defined in the same domain by choosing the complementary vector that fixes $q_A^i$'s labels $q_{A0}^i$ as $v_B^i = v_C^i - v_A^i$ and applying formula (4.2). ∎

It is this implication of the theorem that we shall employ to connect different trajectory models of the quantum state. Whether there are general computational advantages in obtaining the integral curves of a vector $v_C^i$ through an adroit choice of $q_A^i$ coupled with the calculation of the integral curves of $V_B^i$ merits further investigation.

Our result formalizes the rather obvious property that each point on a curve in a region lies on an element of some region-filling congruence. This observation is not bereft of physical interest. It is pertinent to a conception of motion that treats a path $q_C^i$ as a series of points lying on paths from which it is continually displaced by some force (whose nature depends on the context). The construction is reminiscent of, and generalizes to arbitrary motions, Newton's conception of the Moon's orbit as comprising a succession of points each of which lies instantaneously on a distinct rectilinear path, the agent that continually displaces the mass from one rectilinear path to another being gravity. In the context of (4.2), an example where forced and force-free motion are connected is obtained by choosing $v_A^i$ = constant.

## 4.2 Conservation from non-conservation

As an application of the corollary, we shall show how the representation of one species of trajectory by another enables the local conservation of probability along a track (such as a de Broglie-Bohm $q^i$) to be derived from a flow (such as $q_+^i$) along each of whose tracks the probability is not conserved.

Let $v_C^i$ be associated with a current that obeys the continuity equation in some domain (this could be the de Broglie-Bohm velocity, but that choice is not unique [31]):

$$\frac{\partial \rho}{\partial t} + \partial_i(\rho v_C^i) = 0. \qquad (4.8)$$

Writing $v_C^i = v_A^i + v_B^i$, where $v_A^i$ is any vector field in the domain, (4.8) can be written

$$\frac{\partial \rho}{\partial t} + \partial_i(\rho v_A^i) = -\partial_i(\rho v_B^i). \tag{4.9}$$

With respect to the flow $v_A^i$ this is a continuity equation with a 'source' term on the right-hand side. Staying for the moment with general vector fields, we shall derive two key results from (4.9) by transforming it into Lagrangian variables via the mapping $x^i \to q_{A0}^i$. The theory will then be applied to the bi-HJ model, below and in §5. Using the relations

$$\frac{\partial}{\partial q_A^i} = J_A^{-1} J_{Ai}^j \frac{\partial}{\partial q_{A0}^j}, \qquad J_{Ai}^j = \frac{\partial J_A}{\partial(\partial q_A^i/\partial q_{A0}^j)}, \tag{4.10}$$

with $J_A(q_{A0}, t) = \det(\partial q_A/\partial q_{A0})$, and applying the identities $\partial \log J_A/\partial t = \partial \dot{q}_A^i/\partial q_A^i$ and $\partial J_{Ai}^j/\partial q_{A0}^j = 0$, it is readily verified that (4.9) can be written

$$\frac{\partial}{\partial t}(P_A J_A)\bigg|_{q_{A0}} + \frac{\partial}{\partial q_{A0}^i}(P_A J_A V_B^i) = 0 \tag{4.11}$$

where $P_A(q_{A0}, t) = \rho(x = q_A(q_{A0}, t), t)$ and $V_B^i$ is given in (4.1).

Our first deduction from (4.11) stems from the obvious property that the local probability $\rho d^3x$ is not conserved along $v_A^i$'s integral curves. Thus, multiplying (4.11) by $d^3 q_{A0}$, the probability $P_A d^3 q_A(t) = P_A J_A d^3 q_{A0}$ is not generally maintained along a track $q_{A0}^i$, due to the divergence term. This result may be recast as the statement that *the density of the trajectories $q_{A0}^i$ alone does not determine the flow of probability density*. This shows the importance of distinguishing the two notions of density. To show this, we integrate (4.11) with respect to $t$, which puts it into integro-differential form and gives the following expressions for $\rho$:

$$P_A(q_{A0}, t) = \rho_0(q_{A0}) J_A^{-1} + c_A \quad \text{or} \quad \rho(x, t) = (\rho_0 J_A^{-1} + c_A)|_{q_{A0}(x,t)} \tag{4.12}$$

where

$$c_A(q_{A0}, t) = J_A^{-1} \frac{\partial}{\partial q_{A0}^i} \int_0^t P_A J_A V_B^i dt. \tag{4.13}$$

Suppose the trajectories have the (quantum) density $\rho_0$ at $t = 0$. Then the probability density $\rho$ equals the trajectory density for all $t$ if and only if $\rho$ obeys a continuity equation, i.e., the flow is source-free. This follows from (4.12). The first term on the right-hand side, $\rho_0 J_A^{-1}$, represents the density of trajectories at time $t$ due to their natural evolution. This term satisfies the continuity equation corresponding to the velocity $v_A^i$ [7]. Clearly, $\rho$ and $\rho_0 J_A^{-1}$ coincide if and only if the second term, the source $c_A$, is zero, which proves our assertion. In the present case $c_A$, which involves $\rho$, does not vanish. Hence, the trajectory density alone cannot account for $\rho$'s evolution.

---

[7] For a single moving point, the microscopic spatial density at the point $(x^i, t)$ is $\delta(x - q_A(q_{A0}, t))$, which obeys the continuity equation (with velocity $v_A^i$) identically. For a continuous ensemble of points, the (normalized) trajectory density is $\int \delta(x - q_A(q_{A0}, t)) \rho_0(q_{A0}) d^3 q_{A0} = \rho_0 J_A^{-1}|_{q_{A0}(x,t)}$, which, being a linear superposition over the path label, obeys the same continuity equation.

Our second result relates to the corollary. To obtain conservation along a path derived from the flow $v_A^i$, an ensemble of tracks $q_{A0}^i$ selected from the entire associated congruence must be employed. To demonstrate this, we observe that the Lagrangian-picture relation (4.11) has the form of an Eulerian continuity equation in $q_{A0}^i$-space for a flow of density $P_A J_A$ and velocity $V_B^i$. Then, treating (4.11) as such an Eulerian equation, we may pass in turn to *its* Lagrangian version. To effect this passage, we transform from what are now regarded as Eulerian coordinates $q_{A0}^i$ to Lagrangian coordinates $Q_B^i(q_{C0}, t)$, which are the integral curves of the velocity field $V_B^i$ in $q_{A0}^i$-space. In the coordinates $(q_{C0}, t)$, (4.11) becomes

$$\frac{\partial}{\partial t}[(P_A J_A)(q_{A0} = Q_B(q_{C0}, t), t) J_B(q_{C0}, t)]\bigg|_{q_{C0}} = 0 \qquad (4.14)$$

where $J_B(q_{C0}, t) = \det(\partial Q_B/\partial q_{C0})$. Now, $P_A(q_{A0} = Q_B(q_{C0}, t), t) = \rho(x = q_A(q_{A0} = Q_B(q_{C0}, t), t), t)$. Enlisting the theorem in §4.1, the function $q_A^i(Q_B(q_{C0}, t), t)$ in the latter relation is an integral curve $q_C^i(q_{C0}, t)$ of $v_C^i$ with label $q_{C0}^i$. Hence, (4.14) may be written

$$\frac{\partial}{\partial t}(P_C J_C)\bigg|_{q_{C0}} = 0 \qquad (4.15)$$

where $P_C(q_{C0}, t) = \rho(x = q_C(q_{C0}, t), t)$ and $J_C(q_{C0}, t) = \det(\partial q_C/\partial q_{C0}) = J_A J_B$. But (4.15) is just the Lagrangian version of (4.8), which expresses the conservation of the probability $P_C d^3 q_C(t)$ along the integral curves of $v_C^i$, that is,

$$\rho(x, t) = J_C^{-1}(q_{C0}, t)\rho_0(q_{C0})|_{q_{C0}(x,t)}. \qquad (4.16)$$

We conclude that equation (4.9) with a source may be construed as a statement about the conservation of probability along paths that cross the integral curves of $v_A^i$ in a specific way.

Passing now to the bi-HJ theory, we set $v_A^i = v_\pm^i$ and (4.9) becomes a brace of Fokker-Planck equations:

$$\frac{\partial \rho}{\partial t} + \partial_i(\rho v_\pm^i) = \pm\frac{\hbar}{2m}\partial_{ii}\rho. \qquad (4.17)$$

As shown above in connection with (4.12), neither of the congruences' spatial densities, $\rho_0 J_\pm^{-1}$, can account for the flow of $\rho$. Moreover, they cannot play this role in combination. Suppose the initial congruence densities are $r\rho_0$ (+) and $(1-r)\rho_0$ (−), $0 \leq r \leq 1$, so that the initial total density is $\rho_0$. Then the total density for $t \neq 0$ is given by the weighted sum of the expressions (4.12) for $A = \pm$:

$$\rho(x, t) = r(\rho_0 J_+^{-1} + c_+)|_{q_{+0}(x,t)} + (1-r)(\rho_0 J_-^{-1} + c_-)|_{q_{-0}(x,t)}. \qquad (4.18)$$

Evidently, the total trajectory density (the sum of the first and third terms) fails to equal $\rho$ because the $c_+$ and $c_-$ terms do not cancel. As we have seen, the congruences generate $\rho$ instead through the actions accumulated along their elements, in accordance with (2.7).

## 5 The de Broglie-Bohm trajectory as a succession of points on non-T paths, and vice versa

We shall illustrate the corollary (§4.1) by establishing constructive connections between T-complying de Broglie-Bohm trajectories and T-violating trajectories in the case of a one-dimensional free Gaussian wavefunction at rest, for which

$$\rho(x,t) = (2\pi\sigma^2)^{-1/2} e^{-x^2/2\sigma^2}, \quad S(x,t) = \hbar\kappa t x^2/4\sigma^2 - (\hbar/2)\tan^{-1}\kappa t, \qquad (5.1)$$

where $\sigma = \sigma_0(1 + \kappa^2 t^2)^{1/2}$ and $\kappa = \hbar/2m\sigma_0^2$ [9]. The de Broglie-Bohm paths are

$$q(q_0, t) = q_0(1 + \kappa^2 t^2)^{1/2}. \qquad (5.2)$$

The bi-HJ trajectories, solutions to (3.1), are (Fig. 2)

$$q_\pm(q_{\pm 0}, t) = q_{\pm 0}(1 + \kappa^2 t^2)^{1/2} e^{\mp \tan^{-1}\kappa t}. \qquad (5.3)$$

The exponential factors here are suggestive of dissipative or anti-dissipative behaviour, a point we shall pursue elsewhere. The scales and constants in Figs. 2 and 3 are chosen to accentuate the features of interest.

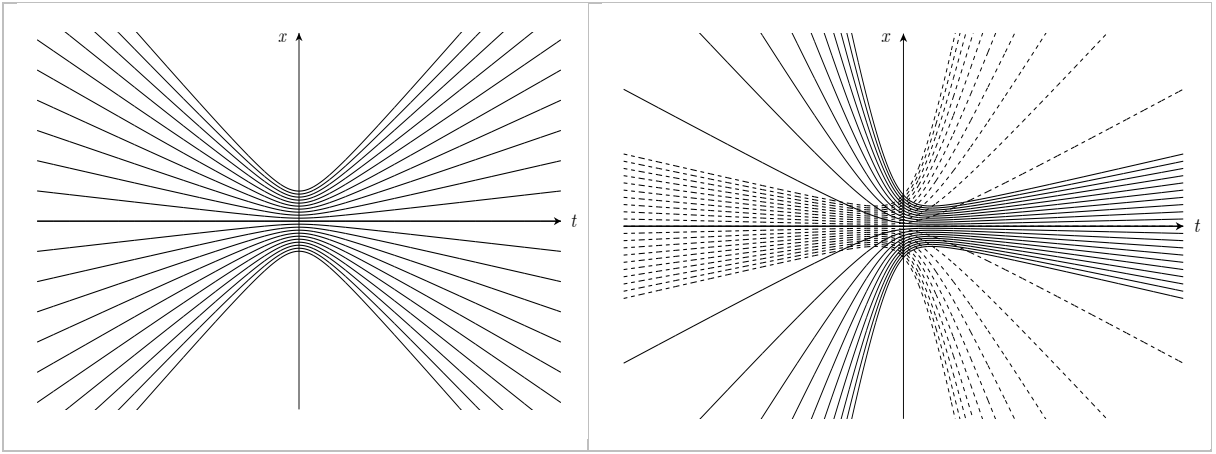

Fig. 2. Two views of diffraction at a Gaussian slit (uniform distributions of initial positions). Left: The state as a de Broglie-Bohm congruence $q(q_0, t)$. The probability density is the normalized path density and the phase is the action. The time-reversed solution is $q(q_0, -t)$. Right: The state as the pair of congruences $q_+(q_{+0}, t)$ (solid lines) and $q_-(q_{-0}, t)$ (dashed lines). The probability density derives from the difference in the actions along the paths and the phase is the mean action. The time-reversed solutions are $q_\mp(q_{\mp 0}, -t)$.

The technique of obtaining a de Broglie-Bohm path from non-T trajectories will be illustrated in two ways, first using HJ paths and then some closely related non-HJ trajectories:

(i) To obtain $q^i$ from $q^i_+$, choose $v^i_A = v^i_+$ and $v^i_B = -\frac{1}{2}u^i = -(\hbar/2m)\partial_i\log\rho$ so that $v^i_C = v^i = \partial_i S/m$. Then $q^i_A = q^i_+$, $q^i_C = q^i$ and

$$V^i_B(q_{+0}, t) = -\frac{1}{2}\frac{\partial q^i_{+0}}{\partial q^j_+} u^j(q_+(q_{+0}, t), t) \tag{5.4}$$

whose integral curves $q^i_{+0} = Q^i_B(q_0, t)$ are given by the solutions of $\dot{Q}^i_B = V^i_B(Q_B, t)$. At an instant $t$, the de Broglie-Bohm path with initial point $q^i_0$ is a point on the track $q^i_+(q_{+0}, t)$ corresponding to the instantaneous label $q^i_{+0} = Q^i_B(q_0, t)$. As the label evolves continuously with time, $q^i$ comprises a succession of points on different $q^i_+$ curves. For the Gaussian, we have $Q_B(q_0, t) = q_0 e^{\tan^{-1}\kappa t}$. Substituting the latter for $q_{+0}$ in (5.3) gives (5.2) (Fig. 3). Note that $q^i$ coincides with some $q^i_+(q_{+0})$ for a finite time when $v^i = v^i_+$, i.e., $\partial_i\rho = 0$, a case where $S$ is classical (since $Q = 0$).

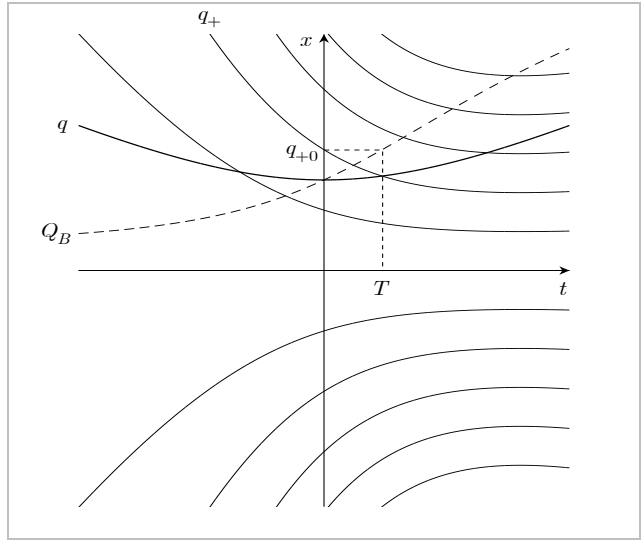

Fig. 3. A de Broglie-Bohm track $q$ (thick line) derived from a sequence of elements of the non-T congruence $q_+(q_{+0})$ (thin lines) for a Gaussian packet. At the instant $T$, the path $Q_B(t)$ (dashed line) identifies, through what was its initial location $q_{+0}$, the trajectory whose instantaneous location $q_+(q_{+0}, T)$ equals $q(q_0, T)$.

(ii) To obtain a de Broglie-Bohm track from a non-T trajectory $q^i_A \neq q^i_+$, choose $v^i_A = \frac{1}{2}v^i_+$ and $v^i_B = \frac{1}{2}v^i_-$, whence $v^i_C = \partial_i S/m$ and $q^i_C = q^i$ as in (i). For the Gaussian, the paths are $q_A = q_{A0}(1 + \kappa^2 t^2)^{1/4} e^{-\tan^{-1}\kappa t/2}$ and $Q_B(q_0, t) = q_0(1 + \kappa^2 t^2)^{1/4} e^{\tan^{-1}\kappa t/2}$. Substituting $Q_B$ for $q_{A0}$ gives (5.2).

To demonstrate the converse procedure, the construction of a non-T curve from a sequence of de Broglie-Bohm paths, we choose $v^i_A = v^i$ and $v^i_B = v^i_+ - v^i = \frac{1}{2}u^i$. Then $q^i_A = q^i$, $q^i_C = q^i_+$ and $V^i_B$ is given by the negative of (5.4). For the Gaussian, $Q_B(q_{+0}, t) = q_{+0} e^{-\tan^{-1}\kappa t}$ so that, substituting $Q_B$ for $q_0$ in (5.2), $q_+$ in (5.3) follows.

# 6 Conclusion

We have presented a novel formulation of quantum evolution based on an extension of the Hamilton-Jacobi method whereby the dynamics is represented by a pair of real HJ-like equations that are coupled by terms we have identified as quantum potentials. The four principal results are as follows:

(i) The quantum state may be taken to comprise two vector displacement functions $q_\pm^i(q_{\pm 0}, t)$. In this setting, the Schrödinger equation becomes an autonomous coalition of equations, (3.3) and (3.4), or (3.7), from which the wavefunction has been eliminated and appears only in the initial conditions. Each pair of trajectories through a spacetime point generates the wavefunction as a pair of action functions $S_\pm$ by propagating $S_{\pm 0}$ from unique initial points $q_{\pm 0}^i$. Varying the latter, the formula (3.11) gives the time-dependent wavefunction throughout space. Each description of state — wave or trajectory — inheres in the other, the connection being mediated by a transformation of independent variables.

(ii) Hitherto, quantum trajectory theories, both constructive and interpretative, have relied upon conserved flows associated with continuity equations obeyed by the probability density $\rho$ or other functions identified as densities (§1). This is not necessary: quantum propagation may be attributed to trajectory flows without reference to conservation, including that of probability. Indeed, we have shown that the conserved flows of the two congruence densities do not suffice to reproduce $\rho$'s development (§4.2). The latter is derived instead from the difference in the actions: $\rho = e^{(S_+ - S_-)/\hbar}$. Its peaks and troughs are characterized not by the relative bunching of trajectories but by points at which the relative velocity $v_+^i - v_-^i = 0$. The mean action is the phase.

(iii) We have illustrated in a new way a feature we found previously in connection with continuous symmetries of quantum field equations: that the Eulerian and Lagrangian elements of a Galileo- or Lorentz-covariant theory may not obey classical transformation rules [5,7,12]. In the context of discrete symmetries of the Galileo-covariant bi-HJ theory, we have found that microscopic time-reversal covariance is implemented through non-standard mappings of the velocity fields (Eulerian view: (2.11)), and of the labels and displacements (Lagrangian view: (3.15)). The theory exhibits effective T-symmetry through the collective properties of non-T elements.

(iv) Using formula (4.2), we have shown how a trajectory, an integral curve of a vector field, may be computed algebraically from the integral curves of another vector field (bestriding the same region) labelled by integral curves associated with the complementary vector. The construction establishes relations between the displacement functions in qualitatively distinct trajectory models, including where the functions have different symmetry properties. The method transcends our quantum examples but finds fecund application there, as illustrated in §5. It also invites further examination of instances where it is fruitful to regard force as an agent of continual deviation from a sequence of 'fiducial' motions, whose choice depends on the context.

In relation to the single task of visualizing quantum evolution using trajectories, the two ways of seeing described in §1 together with the third view introduced in this paper are not contradictory or inimical to objectivity; rather, they extend the transformation theory of quantum mechanics and expand the horizon of how its empirical content, about which they all agree, may be apprehended. In this work, we have focused on formal properties of the bi-HJ trajectory model. Further work is needed to bring out more fully how the technique extends the HJ language, and to examine its efficacy in solving the wave equation. Regarding the latter, it is evident that the bi-HJ paths give more detailed information about the wavefunction than those of de Broglie and Bohm. Consider the class of wavefunctions $\psi_n = \sqrt{\rho_n(x,t)} \exp\{i[f_n(t) + s(x,t)]/\hbar\}$ itemized by the parameter $n$ (e.g., the Gauss-Hermite functions [32]). The de Broglie-Bohm paths are identical for all choices of the probability density $\rho_n$ whereas the bi-HJ tracks depend on $n$ and hence distinguish the densities. A further remark on methods of solution is that one may envisage developing numerical schemes to solve the Eulerian and Lagrangian equations simultaneously, analogous to the synthetic method pioneered by Wyatt and colleagues for the de Broglie-Bohm paths [33]. Since the bi-HJ trajectory densities are not tied to probability, the paths may be useful in exploring domains of low probability, which the de Broglie-Bohm trajectories tend to avoid.

# References

1. Holland, P.: Computing the wavefunction from trajectories: particle and wave pictures in quantum mechanics and their relation. Ann. Phys. (NY) **315**, 505–531 (2005). https://doi:10.1016/j.aop.2004.09.008
2. Holland, P.: Hydrodynamic construction of the electromagnetic field. Proc. R. Soc. A **461**, 3659–3679 (2005). https://doi:10.1098/rspa.2005.1525
3. Holland, P.: Three-dimensional representation of the many-body quantum state. J. Mol. Model. **24**, 269 (2018). https://doi:10.1007/s00894-018-3804-7
4. Holland, P.: The quantum state as spatial displacement. In: Kastner, R.E., Jeknić-Dugić, J., Jaroszkiewicz, G. (eds.) Quantum Structural Studies: Classical Emergence from the Quantum Level, Chap. 10. World Scientific, London (2017).
5. Holland, P.: Schrödinger dynamics as a two-phase conserved flow: an alternative trajectory construction of quantum propagation. J. Phys. A: Math. Theor. **42**, 075307 (2009). https://doi:10.1088/1751-8113/42/7/075307
6. Holland, P.: Trajectory-state theory of the Klein-Gordon field. Eur. Phys. J. Plus **134**, 434 (2019). https://doi.org/10.1140/epjp/i2019-12922-5
7. Holland, P.: Trajectory construction of Dirac evolution. Proc. R. Soc. A.**476,** 20190682 (2020). https://doi: 10.1098/rspa.2019.0682
8. Bennett, A.: Lagrangian Fluid Dynamics. Cambridge University Press, Cambridge (2006)
9. Holland, P.R.: The Quantum Theory of Motion. Cambridge University Press, Cambridge (1993).

https://doi: 10.1017/CBO9780511622687
10. Holland, P.: Hamiltonian theory of wave and particle in quantum mechanics I: Liouville's theorem and the interpretation of the de Broglie-Bohm theory; II: Hamilton-Jacobi theory and particle back-reaction. Nuovo Cimento B **116**, 1043-1069; 1143-1172 (2001).
11. Bokulich, A.: Losing sight of the forest for the psi: beyond the wavefunction hegemony. In: French, S., Saatsi, J. (eds.) Scientific Realism and the Quantum, Chap. 10. Oxford University Press, Oxford (2020)
12. Holland, P.: Hydrodynamics, particle relabelling and relativity. Int. J. Theor. Phys. **51**, 667–683 (2012). https://doi:10.1007/s10773-011-0946-0
13. Holland, P.: Uniting the wave and the particle in quantum mechanics. Quantum Stud.: Math. Found. **7**, 155–178 (2020). https://doi:10.1007/s40509–019–00207–4
14. Holland, P.: Unification of the wave and guidance equations for spin 1/2. Quantum Stud.: Math. Found. **8**, 157-166 (2021). https://doi: 10.1007/s40509-020-00234-6
15. Bowen, R.M.: Theory of mixtures. In: Eringen, A.C. (ed.) Continuum Physics Vol. 3: Mixtures and EM Field Theories, pp. 1-127. Academic Press, New York (1976)
16. Drumheller, D.S., Bedford, A.S.: A thermomechanical theory for reacting immiscible mixtures. Arch. Rational Mech. Anal. **73**, 257-284 (1980). https://doi: 10.1007/BF00282206
17. de Broglie, L.: La Thermodynamique de la Particule Isolée. Gauthier-Villars, Paris (1964)
18. Tolman, R.C.: The Principles of Statistical Mechanics. Dover, New York (1979)
19. Sachs, R.G.: The Physics of Time Reversal. University of Chicago Press, Chicago (1987)
20. Roberts, B.W.: Time reversal. In: Knox, E., Wilson, A. (eds.) The Routledge Companion to Philosophy of Physics, Chap. 43. Routledge, London (2021)
21. Nelson, E.: Quantum Fluctuations. Princeton University Press, Princeton (1985)
22. Cufaro Petroni, N.: Probability and Stochastic Processes for Physicists. Springer, Cham (2020). https://doi: 10.1007/978-3-030-48408-8
23. Leacock, R.A., Padgett, M.S.: Hamilton–Jacobi theory and the quantum action variable. Phys. Rev. Lett. **50**, 3-6 (1983). https://doi: 10.1103/PhysRevLett.50.3
24. Leacock, R.A., Padgett, M.S.: Hamilton–Jacobi/action-angle quantum mechanics. Phys. Rev. D **28**, 2491-2502 (1983). https://doi: 10.1103/PhysRevD.28.2491
25. John, M.V.: Probability and complex quantum trajectories: Finding the missing links. Ann. Phys. (NY) **325**, 2132-2139 (2010). https://doi: 10.1016/j.aop.2010.06.008
26. Chou, C.C., Wyatt, R.E.: Considerations on the probability density in complex space. Phys. Rev. A **78**, 044101 (2008). https://doi: 10.1103/PhysRevA.78.044101
27. Lamb, H.: Hydrodynamics, 6th edn. Cambridge University Press, Cambridge (1932)
28. Truesdell, C.: The Kinematics of Vorticity. Indiana University Press, Bloomington (1954)
29. Blaszak, M.: Multi-Hamiltonian Theory of Dynamical Systems. Springer-Verlag, Berlin (1998)
30. Holland, P.: Dynamics-dependent symmetries in Newtonian mechanics. Phys. Scr. **89**, 015101 (2014). https://doi.org/10.1088/0031-8949/89/01/015101
31. Holland, P., Philippidis, C.: Implications of Lorentz covariance for the guidance equation in two-slit quantum interference. Phys. Rev. A **67**, 062105 (2003). https://doi:10.1103/PhysRevA.67.062105
32. van Dijk, W., Masafumi Toyama, F., Prins, S.J., Spyksma, K.: Analytic time-dependent solutions of the one-dimensional Schrödinger equation. Am. J. Phys. **82**, 955-961 (2014). https://doi:10.1119/1.4885376
33. Wyatt, R.E.: Quantum Dynamics with Trajectories. Springer, New York (2005)